\documentclass{elsart}

\usepackage{graphicx,natbib,amssymb,lineno}

\usepackage{epstopdf}
\usepackage{bm}

\makeatletter
\def\elsartstyle{%
    \def\normalsize{\@setfontsize\normalsize\@xiipt{14.5}}
    \def\small{\@setfontsize\small\@xipt{13.6}}
    \let\footnotesize=\small
    \def\large{\@setfontsize\large\@xivpt{18}}
    \def\Large{\@setfontsize\Large\@xviipt{22}}
    \skip\@mpfootins = 18\p@ \@plus 2\p@
    \normalsize
}
\@ifundefined{square}{}{\let\Box\square}
\makeatother

\begin{document}

\begin{frontmatter}

\title{Bose-Einstein condensates in disordered potentials}
\author{Leonardo Fallani, Chiara Fort, and Massimo Inguscio}
\address{LENS European Laboratory for Nonlinear
Spectroscopy\\and Dipartimento di Fisica, Universit\`a di Firenze\\
Via Nello Carrara 1, 50019 Sesto Fiorentino (FI), Italy}

\begin{abstract}
The interplay between disorder and interactions is a
\emph{leit-motiv} of condensed matter physics, since it constitutes
the driving mechanism of the metal-insulator transition.
Bose-Einstein condensates in optical potentials are proving to be
powerful tools to quantum simulate disordered systems. We will
review the main experimental and theoretical results achieved in the
last few years in this rapidly developing field.
\end{abstract}


\end{frontmatter}

\tableofcontents


\section{Introduction}

In Nature many processes occur in an ordered way. Indeed, ordered
configurations are often the ones minimizing the total energy of the
system. A prominent example of this tendency towards order is given
by the growth of a crystal, where the atoms arrange themselves in a
spatially periodic configuration building up an ordered lattice. The
physics of transport of electrons in a metal heavily relies on the
periodicity of this lattice. However, when crystalline solids are
studied on a sufficiently small length scale, one realizes that
impurities and defects are always present, which may affect in a
substantial way the transport of the electrons. Disorder is indeed
an intrinsic property of all the real systems. In the last 50 years
the effects of disorder on transport phenomena have been extensively
studied in the context of both statistical and condensed matter
physics.

Despite the very general interest in understanding the physics of
disorder in condensed matter systems, still many questions remain
open and unsolved, even from the theoretical point of view. As a
matter of fact, the theoretical description of periodic systems, as
perfect crystals, is much easier than the one for disordered system
as disordered lattices or glasses. The problems that arise are
related to the fact that the effects of disorder cannot be
theoretically treated in a perturbative way: even a small amount of
disorder can produce dramatic changes in the physical properties of
the system under investigation. In 1958, P. W. Anderson published a
seminal paper \citep{anderson58} in which he showed under which
conditions non-interacting electrons in a disordered metal can
either move through the system, or be localized. It was soon
realized that Anderson localization is a much more general
phenomenon holding for the propagation of generic classical waves in
disordered media. Localization is a coherent effect that arises from
multiple scattering of a wave from randomly-distributed impurities
and from the resulting destructive interference in the direction of
propagation.

Also interactions are well known to induce localization effects, as
pointed out by N. F. Mott who was able to explain the anomalous
insulator behavior of some materials when electron-electron
interactions were included in the band theory. In 1977 P. W.
Anderson and N. F. Mott were awarded with the Nobel Prize in Physics
for their fundamental theoretical investigations of the electronic
structure of magnetic and disordered systems
\citep{anderson78,mott78}. Following these pioneering works, a
strong theoretical effort has been devoted in the last decades to
investigate the combined role of disorder and interactions in the
superfluid-insulator transition observed in many condensed-matter
systems, such as $^4$He adsorbed on porous media \citep{crowell95},
thin superconducting films \citep{goldman98}, arrays of Josephson
junctions \citep{vanderzant92} and high-temperature superconductors
\citep{jiang94,budhani94}.

Ultracold atoms in optical lattices \citep{morsch06} represent an
extremely powerful tool for engineering simple quantum systems with
a broad tunability of the parameters, thus serving as ``quantum
simulators" \citep{feynman82} to reproduce the physics of different
systems. The striking advantage offered by such atomic systems
resides in the unprecedented possibility to work with perfectly
isolated samples at quasi-zero temperature and to have experimental
control on most of the Hamiltonian parameters, e.g. the lattice
depth or the strength of the atom-atom interactions, that can be
precisely tuned even in real-time. One spectacular demonstration of
this opportunity has been given by the observation of the superfluid
(SF) to Mott insulator (MI) transition in a 3D optical lattice
\citep{greiner02}, which pioneered the investigation of strongly
quantum correlated regimes with ultracold atoms \citep{bloch07}.

A natural extension of these experiments is the realization of
disordered systems using ultracold atoms in optical potentials. In
this paper we will review the recent progresses in this field, that
was experimentally initiated in 2004 with the first investigation of
atomic Bose-Einstein condensates in disordered potentials. Different
possibilities can be followed to produce disordered ultracold atomic
systems. Disordered or quasi-disordered potentials can be created
optically by using speckle patterns \citep{lye05} or multi-chromatic
incommensurate optical lattices \citep{fallani07}. These methods
allow the production of disordered potentials in which both the
spectral properties and the amount of disorder are known with very
good accuracy and can be easily controlled. In addition to the
optical way, disordered systems could also be created by using
atomic mixtures \citep{gavish05} or inhomogeneous magnetic fields
\citep{gimperlein05,fortagh06}. We will review these different
possibilities together with the illustration of the diverse
interaction regimes that can be investigated. The first experimental
results obtained with ultracold bosons in disordered potentials will
be presented, discussing the state of the art of this newborn field
and the perspectives for future breakthroughs.


\section{How to produce a disordered potential}

In this section we will present different experimental approaches to
the production of disordered potentials for neutral atoms. We will
mostly focus on two methods allowing the production of complex
optical potentials: speckles patterns and multichromatic lattices.

\subsection{Speckle patterns}

\begin{figure}
\begin{center}
\includegraphics[width=\columnwidth]{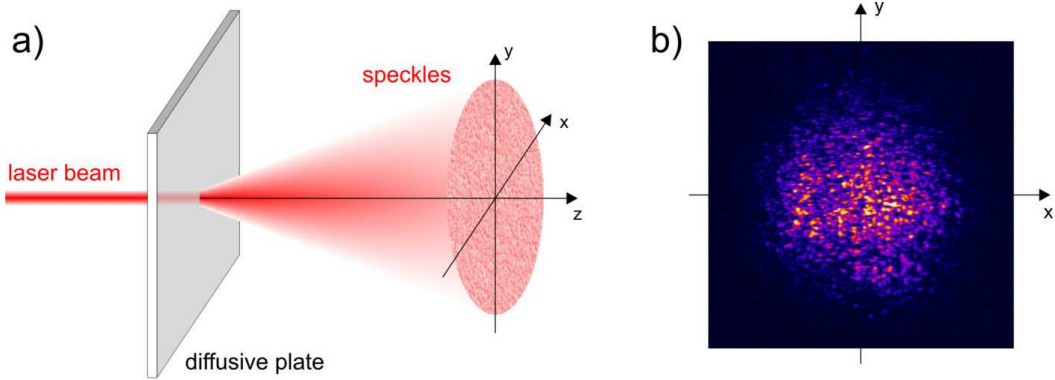}
\end{center}
\caption{Production of speckle patterns. a) A laser beam is shone
through a diffusive plate and the resulting speckle pattern is then
imaged onto the BEC. b) Intensity distribution of a typical speckle
pattern recorded with a CCD camera.} \label{fig:specklesetup}
\end{figure}

The first realization of disordered potentials for cold atoms has
been obtained with speckle patterns \citep{boiron99}. Speckles are
produced whenever light is reflected by a rough surface or
transmitted by a diffusive medium \citep{goodman07}. We will mostly
consider the case of transmission, sketched in Fig.
\ref{fig:specklesetup}a, and we will refer to the scattering device
as a \emph{diffusive plate}. Such device can be modeled as made up
of many randomly-distributed impurities by which the illuminating
light is scattered. Since the scattering of laser light is mainly a
coherent process, the partial waves emerging from the scattering
interfere and produce a complex distribution of light, called a
\emph{speckle pattern}, an example of which is shown in Fig.
\ref{fig:specklesetup}b. This disordered distribution of light can
be imaged onto the atoms, producing a disordered potential
$V(\mathbf{r})$ proportional to the local laser intensity
$I(\mathbf{r})$.

In general, if the wavelength of the light is far detuned from the
atomic resonance, no absorption is involved and the resulting
mechanical effect can be described by a potential energy of the form
\begin{equation}
V(\mathbf{r}) = \frac{3 \pi c^2}{2 \omega_0^3} \left(
\frac{\Gamma}{\Delta}\right) I(\mathbf{r}) \; , \label{eq:potdip}
\end{equation}
where $c$ is the speed of light, $\omega_0$ is the frequency of the
atomic resonance, $\Gamma$ its radiative linewidth, $\Delta =
\omega-\omega_0$ the detuning, and $I(\mathbf{r})$ the intensity
distribution. This potential is often called \textit{dipole
potential} \citep{grimm00}. It is worth noting that the sign of this
potential depends on $\Delta$, which is the only quantity which can
take either positive or negative values. In particular, when
$\Delta<0$ (\textit{red detuning}) $V(\mathbf{r})$ is negative,
hence maxima of light intensity correspond to potential minima:
atoms will move towards higher-intensity regions. Instead, when
$\Delta>0$ (\textit{blue detuning}) $V(\mathbf{r})$ is positive,
hence maxima of light intensity correspond to potential maxima:
atoms will move towards lower-intensity regions.

\begin{figure}
\begin{center}
\includegraphics[width=\columnwidth]{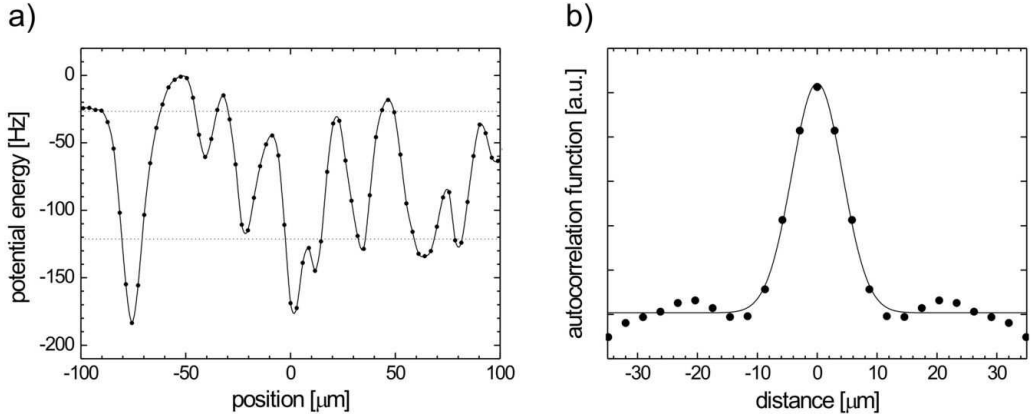}
\end{center}
\caption{Production of speckle patterns. a) Cross section of a
typical speckle potential (the energy difference between the
horizontal lines is the average speckle height $V_S$). b)
Autocorrelation integral of the same potential.}
\label{fig:specklesetup2}
\end{figure}

Speckle patterns represent a valuable way to produce a disordered
potential in a controlled way. The possibility to accurately measure
the statistical and correlation properties of the disordered
potential comes from the fact that the intensity of the speckle
pattern can be directly recorded by a CCD camera (typically the same
one used to image the BEC atoms). In Fig. \ref{fig:specklesetup2}a
we show the cross section of a typical speckle pattern used at LENS
for the first investigation of disordered Bose-Einstein condensates
\citep{lye05}. Among the different quantities characterizing the
properties of the speckle field, one can define an \emph{average
speckle height} $V_S$. Different definitions are used in literature,
however one of the most used in the context of BEC experiments
corresponds to taking twice the standard deviation of the speckle
potential $V(x)$ (supposed one-dimensional) around its mean value
$\overline{V}$ \citep{lye05}:
\begin{equation}
V_S = 2 \left[ \frac{1}{L} \int ^{L/2}_{-L/2} \left(
V(x)-\overline{V} \right) ^2 dx \right] ^ {1/2} \; .
\end{equation}
An even more important quantity, as we shall see in the following,
is the \emph{autocorrelation length} $\sigma$, giving information on
the speckle grain size \citep{goodman07}. This quantity is defined
as the rms width of the autocorrelation integral $G(d)$ of the
speckle potential
\begin{equation}
G(d) = \int ^{L/2}_{-L/2} V(x) V(x+d) dx \approx
e^{-\frac{d^2}{2\sigma^2}} \; ,
\end{equation}
an example of which is shown in Fig. \ref{fig:specklesetup2}b. The
autocorrelation length $\sigma$ depends on the wavelength of the
light, on the nature of the diffusive medium producing the speckle
pattern and, most importantly, on the optical resolution of the lens
system used to image the speckle pattern onto the atoms. As a matter
of fact, the typically determined autocorrelation length is set by
the diffraction limit spot size of the imaging system. A detailed
description of the speckle potential is given in
\citet{clement05njp}, where the statistical properties of the
speckle field are discussed both from a theoretical point of view
and with the introduction of experimental methods which allow their
precise determination.

The random potential produced by a speckle pattern is static. This
means that the atoms experience just one realization of disorder,
which can be reproduced in the same way from one experiment to
another. However, shifting the position of the diffuser leads to a
different realization of the speckle pattern that preserves the same
spectral and statistical properties. Thus, averages on multiple
realizations of disorder can be achieved in a simple way.

Before concluding this section, we note that speckles are
intrinsically two-dimensional in the plane perpendicular to the
propagation axis. Actually, a speckle pattern also varies along the
direction of propagation of the light. However, the typical
correlation size along this direction is much larger. Nonetheless,
speckle potentials with different dimensionality can be produced: 1D
speckles can be produced by using cylindrical lenses stretching the
speckle pattern along one direction, while 3D speckles could be
obtained by adding speckle patterns coming from different
directions.

\subsection{Multichromatic lattices}
\label{sec:bichromatic}

As we have seen in the previous section, speckle patterns are a
powerful and easy-to-implement method to produce random potentials.
We have pointed out that a crucial parameter of such potentials is
the autocorrelation length $\sigma$, which gives an estimate of the
minimum length scale below which the potential loses its random
nature and becomes correlated. Typically, this length is connected
with the diffraction limit dimension at which optical speckles are
imaged onto the atomic sample. For this reason, the random potential
produced by speckles is often too coarse-grained (with $\sigma$ of
the order of several microns), unless one builds a dedicated setup
to overcome the usual optical access restrictions. Recently,
progresses in the realization of speckle potentials with
autocorrelation length below 1 $\mu$m have been achieved in the
experimental groups of A.~Aspect \citep{clement05njp} and
\citet{demarco07}.

On the other side, having in mind many years of exciting physics
with cold atoms in optical lattices, we know that optical standing
waves can be easily created providing spatial periodicities that can
also be smaller than half a micron (roughly speaking, one order of
magnitude smaller than the autocorrelation length of the speckle
pattern shown in Fig. \ref{fig:specklesetup2}). This suggests the
idea that, by combining several optical standing waves with
different non-commensurate spacings, it is possible to produce
complex potentials with very small ``grain size".

\begin{figure}
\begin{center}
\includegraphics[width=0.8\columnwidth]{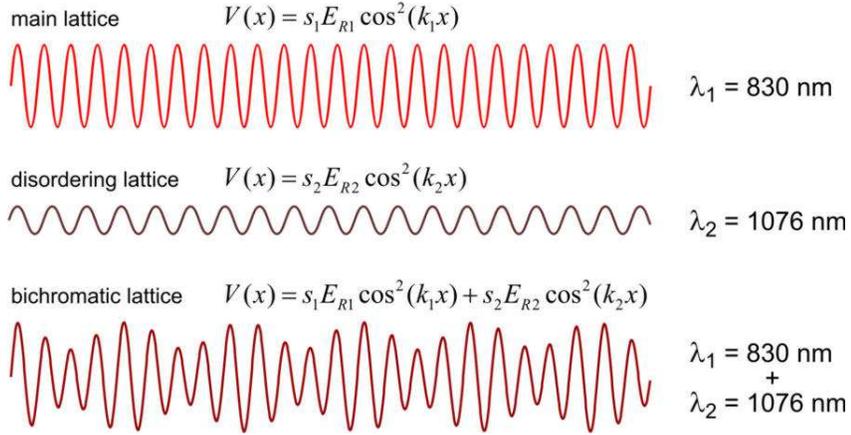}
\end{center}
\caption{A bichromatic optical lattice. The discrete translational
invariance of the main lattice is perturbed by the addition of a
secondary lattice with incommensurate wavelength.}
\label{fig:bichrome}
\end{figure}

The simplest example is given by a bichromatic lattice resulting
from the addition of two lattices with incommensurate wavelengths.
Bose-Einstein condensates in potentials of this kind have been first
investigated in experiments at LENS \citep{fallani07}, where a main
lattice with wavelength $\lambda_1$ was perturbed by a weaker
secondary lattice with wavelength $\lambda_2$, as sketched in Fig.
\ref{fig:bichrome}. The resulting potential can be written in the
form
\begin{equation}
V(x)=s_1 E_{R1} \cos^2 \left( k_1 x \right) + s_2 E_{R2} \cos^2
\left( k_2 x \right)
\end{equation}
where $k_1=2\pi/\lambda_1$ and $k_2=2\pi/\lambda_2$ are the lattice
wavenumbers and $s_1$ and $s_2$ are adimensional numbers indicating
the heights of the two lattices in units of the recoil energies
$E_{R1}=h^2/(2 m \lambda_1^2)$ and $E_{R2}=h^2/(2 m \lambda_2^2)$,
respectively. In the limit $s_2\ll s_1$ the height of the optical
barriers is roughly constant across the whole lattice and it is
possible to define a tunneling rate $J$ which only depends on the
main lattice height $s_1$. In this limit, the effect of the
secondary lattice reduces to an inhomogeneous and non-periodic shift
of the potential energy at the bottom of the lattice wells (see Sec.
\ref{sec:weaklybichromatic} and Sec. \ref{sec:stronglyinteracting}).

As we will further discuss in the following, bichromatic
incommensurate lattices are not truly disordered potentials. They
differ from both purely random potentials and speckle potentials,
which exhibit different statistical and correlation properties.
Strictly speaking, they are \emph{quasiperiodic} potentials, since
their spectrum is made up of a set of discrete frequencies. However,
because of the lack of any translational invariance, they can be
used to investigate the physics of finite-sized disordered systems
and study the emergence of quantum localization effects, as we shall
see in Sec. \ref{sec:weaklybichromatic}.

We note that, since one always deals with finite-sized atomic
samples, the notion of incommensurability (i.e. the wavelength ratio
being an irrational number) is a rather sophistic concept, and
should be substituted with a more practical definition. From an
experimental point of view, since the lattice wavelengths are known
with finite precision, the measurement of the ratio
$\lambda_2/\lambda_1$ always gives a rational number. From a
theoretical point of view, it is important to consider that the
finite size of the systems under investigation releases the
constraints on the incommensurability: even a periodic potential
(resulting from a commensurate ratio) does not show any periodicity
if the system size is smaller than the period. The bichromatic
lattice is thus \emph{effectively} incommensurate provided that the
ratio between the wavelengths is far from a ratio between
\textit{simple} integer numbers. More precisely, a bichromatic
lattice can be considered incommensurate whenever the resulting
periodicity (if any) is larger than the system size.

\subsection{Other methods}
\label{sec:otherways}

\begin{figure}
\begin{center}
\includegraphics[width=0.85\columnwidth]{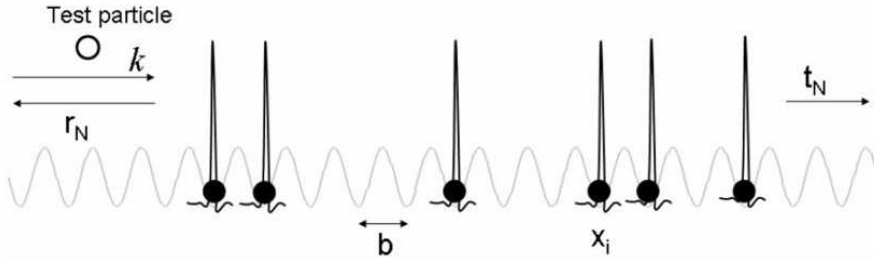}
\end{center}
\caption{Disordered potential produced by the interaction of the
atoms with a different species. An incoming test particle (white)
experiences the collisional potential produced by randomly
distributed (black) atoms trapped in an optical lattice. Taken from
\citet{gavish05}.} \label{fig:castin}
\end{figure}

In solids disorder is often caused by the presence of impurities,
i.e. atoms of a different kind that randomly occupy the sites of the
crystalline lattice where atoms of different species were expected.
This kind of disorder can be simulated also in cold gases by using a
mixture of two different atomic species, as originally proposed in
\citet{gavish05}. The atoms of one of the two species are trapped in
the sites of a deep optical lattice. If the filling factor (i.e. the
average number of atoms per site) is less than unity, only some of
the sites will be occupied by one atom and the other ones will be
empty, as schematically shown in Fig. \ref{fig:castin}. The atoms of
the other species, that could be weakly affected by the presence of
the lattice, feel the collisional interaction with the
randomly-distributed atoms of the first species. This kind of
disorder is spectrally different from both the speckle and
bichromatic potentials, since it is a binary kind of disorder
(yes/no) on top of a periodic backbone. In \citet{gavish05} the
authors have theoretically investigated the possibility to study 1D
Anderson localization of matter waves with this system and this work
has then been extended in 3D in \citet{massignan06}. This scheme has
not been experimentally realized yet, although first experiments
with binary ultracold mixtures have been performed (see Sec.
\ref{sec:atomicmixtures}).

Another way to introduce disorder in the system has been proposed in
\citet{gimperlein05} by using inhomogeneous magnetic fields, e.g. by
exploiting the magnetic field fluctuations in the proximity of a
microtrap caused by imperfections in the chip fabrication
\citep{wang04}. If the bias magnetic field is kept close to a
Feshbach resonance \citep{inouye98}, small field fluctuations on top
of it produce spatial fluctuations in the scattering length
characterizing the interactions between the atoms. Therefore, this
technique allows to introduce disorder on the atom-atom interaction
strength, rather than on the external potential. In
\citet{gimperlein05} the phase diagram of interacting bosons in the
presence of such disorder has been derived, evidencing novel
features with respect to the phase diagram with disorder in the
external potential (that will be presented in Sec.
\ref{sec:stronglyinteractingintro}).


\section{Weakly interacting regime}
\label{sec:weaklyinteractintro}

One of the most fascinating phenomena characterizing the transport
of waves in random systems is \emph{Anderson localization}. This
effect takes its name after the seminal work of P. W. Anderson in
1958, who identified the fundamental role of disorder in the
metal-insulator transition observed in solid state systems
\citep{anderson58}. Anderson first formulated his localization
theory for a simple model of particles hopping on a lattice with
random on-site energies, arguing that above a critical disorder
amplitude the quantum states had to change from extended to
spatially localized. This intuition, together with the mathematical
tools developed to describe the localization transition, led to the
award of the Nobel Prize in Physics in 1977 \citep{anderson78}.

In the following decades, however, it was realized that Anderson
localization is a much more general phenomenon, holding for
propagation of generic linear waves in disordered media. Indeed, it
has been observed for sound waves and light waves
\citep{wiersma97,schwartz07}, whereas a direct observation for
matter waves has not yet been possible.

In the language of wave propagation, Anderson localization arises
because of interference effects in the scattering of a wave by
disordered defects. When studying wave propagation in disordered
systems, different localization regimes can be identified. A
precursor effect of Anderson localization is \emph{weak
localization}, which arises from interference effects in multiple
scattering events: an example of weak localization is given by
\emph{coherent backscattering}, i.e. the enhanced probability of
backdiffusion for light incident on a disordered sample, owing to
the interference between the forward and backward scattering paths
\citep{wiersma95}. In the strong scattering limit $kl \simeq 1$
(with $k$ wavevector and $l$ mean free path between scattering
events), these interferences can add up to completely halt the waves
inside the random medium, resulting in \emph{strong localization},
or Anderson localization.

Anderson localized states are characterized by the typical
exponential decay of their tails in the space distribution. In the
case of light, this means that the intensity is an exponentially
decreasing function of the distance travelled in the disordered
medium. For quantum-mechanical wavefunctions, this means that a
localized state $\Psi(x)$ can be written as
\begin{equation}
\Psi(x) \sim \exp\left(-\frac{x}{\zeta}\right) \; ,
\label{eq:exploc}
\end{equation}
where $\zeta$ is the \emph{localization length}. Generally speaking,
the stronger is the disorder the smaller is the localization length.

In the physics of Anderson localization an important role is played
by the dimensionality of the system. After the first Anderson's
conjecture, scaling arguments have been proposed which predict
different scenarios with changing dimensionality $d$ of the system
\citep{abrahams79}. For $d<2$ all the states are localized. For
$d>2$ a localization transition exists, with a mobility edge
separating extended states for weak disorder from localized states
above a critical value. The 2D case is marginal, since the states
are localized for any amount of disorder as in 1D, but the
localization length at weak disorder can be exponentially large.

A Bose-Einstein condensate is characterized by long-range coherence
and can be described with a classical order parameter which
corresponds to the wavefunction of the Bose-condensed atoms. In the
noninteracting case all the atoms are described by the same
single-particle wavefunction which obeys the Schr\"{o}dinger
equation
\begin{equation}
i \hbar \frac{d\Psi}{dt} = -\frac{\hbar^2}{2m}\nabla^2 \Psi +
V(\mathbf{r})\Psi \; . \label{eq:se}
\end{equation}
In the presence of disorder, this wavefunction can be Anderson
localized. Bose-Einstein condensates represent an appealing system
where it is possible to directly study the effect of localization.
By using the techniques described in the previous section, one is
able to create disordered potentials in an extremely controlled way,
knowing precisely the kind and amount of disorder. Furthermore, the
wavefunction (more precisely, the squared modulus of it) can be
directly observed by imaging the condensate with a CCD camera. As a
result, the typical exponential tails of Anderson localized states
could be observed (at least in principle, if the imaging resolution
and sensitivity are good enough), allowing the detection of
localization.

\begin{figure}[t!]
\begin{center}
\includegraphics[width=\columnwidth]{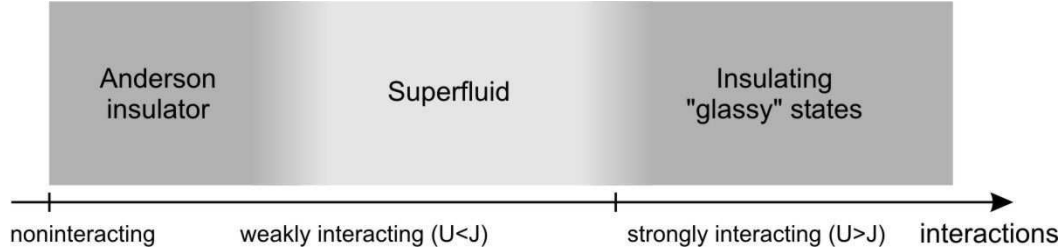}
\end{center}
\caption{Pictorial representation of the different interaction
regimes for a bosonic gas in the presence of disorder.}
\label{fig:sketch}
\end{figure}

An extremely interesting, and still open, problem regards the effect
of interactions on localization. Originally, Anderson formulated his
theory for noninteracting quantum particles. If one considers real
interacting particles, however, the scenario could be significantly
different. In the case of electrons, repulsive long-range
interactions are present due to the Coulomb electric force. In the
case of ultracold neutral atoms long-range dipolar interactions can
be nearly always neglected and the dominant interaction mechanism is
represented by \emph{s}-wave elastic collisions. These short-range
interactions can be either attractive or repulsive, although only
repulsive interactions allow for the existence of stable BECs with
arbitrarily large number of atoms \citep{dalfovo99}. When the BEC
density is sufficiently small, as it happens in many experimental
situations, the effect of the collisional forces can be described
within a mean-field approach by adding a nonlinear term in Eq.
(\ref{eq:se}), which becomes the well-known Gross-Pitaevskii
equation (GPE) \citep{dalfovo99}:
\begin{equation}
i \hbar \frac{d\Psi}{dt} = -\frac{\hbar^2}{2m}\nabla^2 \Psi +
V(\mathbf{r})\Psi + g | \Psi |^2 \Psi \; . \label{eq:gpe}
\end{equation}
The interaction strength, described by $g=4 \pi \hbar^2 a/m$, is
parametrized as a function of one single scalar parameter $a$, which
takes the name of \textit{scattering length}. A similar kind of
cubic nonlinearity is present also in the Maxwell equations
describing the propagation of light in a nonlinear optical medium
where the index of refraction depends on the light intensity (Kerr
effect). This term is responsible for many effects of nonlinear
dynamics, such as solitonic propagation \citep{burger99}, four-wave
mixing \citep{deng99} and instabilities \citep{wu01,fallani04}. The
presence of interactions can heavily affect the physics of
localization, which is intrinsically a single-particle effect,
holding for linear waves. From a naive point of view, negative
nonlinearities ($a<0$, arising from attractive interactions between
particles, or self-focusing behavior of the wave) could play in
favor of localization. On the contrary, positive nonlinearities
($a>0$, induced by repulsive interactions, or self-defocusing
behavior) are expected to play against localization, making the
problem much more interesting to study, both theoretically and
experimentally.

The interplay between disorder and interactions in the physics of
localization has been the object of a very intense theoretical
investigation. It was soon realized that repulsive interactions can
compete with disorder and eventually destroy the localization. In
strongly interacting systems, however, different regimes can be
achieved and new quantum phases can be reached in which interactions
and disorder co-operate in localizing the system in glassy states
(see Sec. \ref{sec:stronglyinteracting} for a discussion of these
phases). An oversimplified picture of the different regimes for an
interacting bosonic gas in a disordered potential is sketched in
Fig. \ref{fig:sketch}. Real phase diagrams, of course, are much more
complicated than this pictorial representation, the details
depending e.g. on the kind of disorder and on the dimensionality of
the system. Actually, there are still many open questions to be
solved, which the experimental study of interacting Bose-Einstein
condensates in disordered potentials could address. As a matter of
fact, these systems offer the advantage of a broad tunability of the
Hamiltonian parameters, including the kind and amount of the
disorder (as seen in the previous section) and the interaction
strength between the atoms (e.g. by using Feshbach resonances
\citep{inouye98}).

\subsection{A Bose-Einstein condensate in a disordered potential}

\subsubsection{Static properties}

A natural starting point to gather information on the behavior of
the BEC in the disordered potential is the shape of the atomic
density distribution after release from the confining potential.
This time-of-flight detection technique has been used since the
first experimental realization of BEC \citep{anderson95} as a
precious tool to study its ground state properties. We start
considering the case of disordered potentials created with optical
speckles, first investigated with $^{87}$Rb in \citet{lye05}.

\begin{figure}[t!]
\begin{center}
\includegraphics[width=0.95\columnwidth]{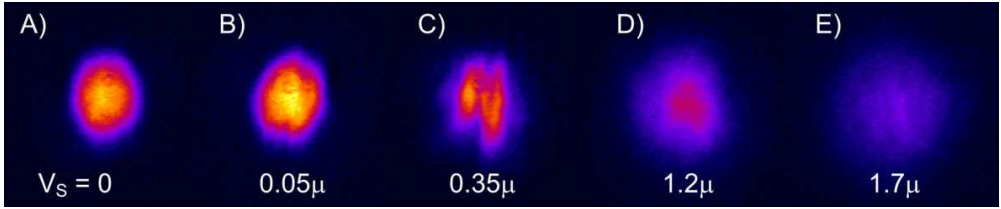}
\end{center}
\caption{Ground state of a BEC in a harmonic + disordered potential.
Absorption images of the atomic density distribution following a
time-of-flight after the release from the confining potential. The
numbers in the bottom indicate the average speckle height $V_S$ in
units of the BEC chemical potential $\mu \simeq 1$ kHz in the
harmonic trap. Adapted from \citet{lye05}.}
\label{fig:randomfringes}
\end{figure}

Basically, one can observe three different regimes, depending on the
ratio between the speckle height $V_S$ and the BEC chemical
potential $\mu$. For very small optical potentials $V_S \lesssim 0.1
\mu$ one does not observe any significant deviation from the
ordinary Thomas-Fermi shape of the BEC expanding from the harmonic
trap (fig. \ref{fig:randomfringes}A,B). For higher speckle heights
$0.1 \mu \lesssim V_S \lesssim \mu$ one observes that the density
distribution is strongly modified by the appearance of complex
structures in the form of elongated stripes (fig.
\ref{fig:randomfringes}C). Finally, further increasing the speckle
height to $V_S \gtrsim \mu$, the expanded density profile ceases to
be characterized by stripes and one can detect only a broad
unstructured gaussian distribution (Fig.
\ref{fig:randomfringes}D,E).

\begin{figure}[t!]
\begin{center}
\includegraphics[width=0.5\columnwidth]{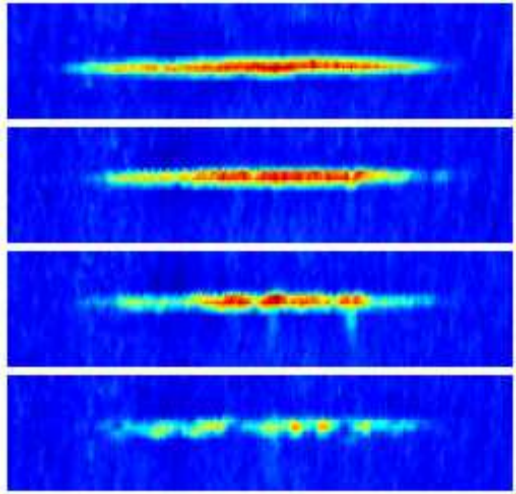}
\end{center}
\caption{Ground state of a BEC in a harmonic + disordered potential.
\emph{In-situ }images of the atomic density distribution. The
speckle potential $V_S$ is increasing from top to bottom: 0,
0.3$\mu$, 0.5$\mu$, 1.3$\mu$. Taken from \citet{chen07}.}
\label{fig:intraphulet}
\end{figure}

The appearance of density modulations in the regime of weak disorder
has recently attracted a large interest. The structures observed
after expansion could arise either from real in-trap density
modulations or from phase fluctuations converted into density
modulations after time-of-flight. More recent experimental works
\citep{chen07,clement07} systematically investigated
such structures, measuring the fringes visibility as a function of
the speckle height and evidencing that the fringes pattern is stable
for the same realization of disorder. This observation, together
with the comparison with GPE simulations \citep{clement07}, suggests
that for the actual experimental parameters the most plausible
scenario is the one in which small in-trap density fluctuations are
amplified during the time-of-flight. The problem is studied in
detail in \citet{clement07} with a thorough analysis of the
mechanisms involved during the BEC expansion.

Further increasing the intensity of the disordered potential to $V_S
\gtrsim \mu$ the condensate is split up in many condensates
localized in the randomly-spaced minima of the speckle potential.
The absence of interference structures in the observed density
distribution is due to the fact that the spacing between different
condensates is not uniform and gives rise to an interference pattern
that, averaged over the optical resolution of the system, is almost
flat. This \emph{fragmentation} scenario has been confirmed in
\citet{chen07} by using direct \emph{in-situ} imaging of the trapped
$^7$Li atoms (see Fig. \ref{fig:intraphulet}).

\subsubsection{Collective excitations}

Low-energy collective excitations of the BEC have been studied in
the presence of a weak speckle disorder producing a corrugation of
the harmonic trap potential. Since the first production of BECs,
frequency measurements of collective modes have provided precious
information for the identification of superfluidity and, more in
general, for the characterization of quantum fluids. In
\citet{lye05} the frequency and the damping rate of the dipole and
quadrupole modes of a BEC in a speckle potential have been measured
as a function of the strength of disorder. With the term of ``dipole
mode" one usually refers to rigid center-of-mass oscillations of the
BEC in the parabolic trap, while the ``quadrupole mode" for a highly
anisotropic trap (or ``axial breathing mode") indicates a shape
oscillation in which the center-of-mass does not move, but the size
of the BEC along its long axis is periodically changing in time
\citep{dalfovo99}. In \citet{lye05} a damping of both modes has been
observed and measured as a function of the strength of the speckle
potential.

Damping of the dipole mode has been recently observed also in
\citet{chen07}, in which the study of the different dynamic regimes
have been accompanied by \textit{in-situ} detection of the atomic
density distribution. Indeed, \citet{chen07} demonstrated that the
halting of the center-of-mass motion for strong disorder has to be
connected with the creation of a fragmented BEC. In this regime
different condensates are trapped in the different speckle potential
wells and no global phase coherence is present, due to the extremely
long tunneling times between different fragments.

\citet{lye05} measured also a frequency shift of the quadrupole mode.
The frequency of the quadrupole mode is particularly
important since its value depends not only on the strength of the
parabolic potential but also on the Bose-Einstein equation of state,
hence it depends on the nature of the system, whether it is
superfluid or not. The frequency changes measured in \citet{lye05},
however, just reflected the change in the effective potential
curvature induced by the corrugation produced by the disordered
potential. The problem has been addressed theoretically in
\citet{modugno06} by numerical solution of the Gross-Pitaevskii
equation combined with a sum-rules approach, confirming the
frequency shift as an effect due to the change in the effective trap
frequency.

The effect of the speckle potential produced in these experiments is
mostly classical, and it does not really produce a change in the
nature of the quantum fluid. We shall discuss more about this point
in the next section, evidencing how the correlation length of the
potential plays a crucial role for the observation of truly
disordered-induced localization effects. In particular, concerning
the measurements of collective excitations, the presence of disorder
with short correlation length can modify the superfluid equation of
state leading to non-trivial frequency shifts, as recently studied
in \citet{falco07}.

\subsection{The quest for Anderson localization}

In this section we will discuss the state-of-the-art of the
experiments aiming to observe Anderson-like localization for
Bose-Einstein condensates propagating in disordered optical
potentials.

\subsubsection{Localization in a speckle potential}

Out of the condensed matter systems for which it has been originally
proposed, Anderson localization has been widely searched, and
eventually demonstrated, in classical wave propagation experiments
\citep{wiersma97,schwartz07}. In this kind of experiments, an
electromagnetic wave undergoes multiple scattering from the
randomly-distributed scatterers of the disordered medium. Strong
(Anderson) localization sets in when the multiple scattered waves
interfere destructively in the propagation direction and localized
states become populated. According to the Ioffe-Regel criterion
\citep{ioffe60} this happens when the mean free path of the wave
becomes as small as its wavelength.

Experiments performed in 2005 at LENS (Florence) and in the group of
A. Aspect at Institut d'Optique (Orsay) aimed to realize such
scattering configuration with Bose-Einstein condensates propagating
in disordered optical potentials produced with speckle patterns
\citep{clement05prl,fort05}. The idea behind these two works was
quite similar: an initially trapped Bose-Einstein condensate of
$^{87}$Rb was left free to expand in a one-dimensional disordered
waveguide. In \citet{clement05prl} this waveguide was produced by a
highly elongated magnetic trap, while in \citet{fort05}  by a single beam
optical trap. The propagation of the
condensed matter wave in the waveguide was studied as a function of
the height of the disordered speckle potential.

\begin{figure}
\begin{center}
\includegraphics[width=0.65\columnwidth]{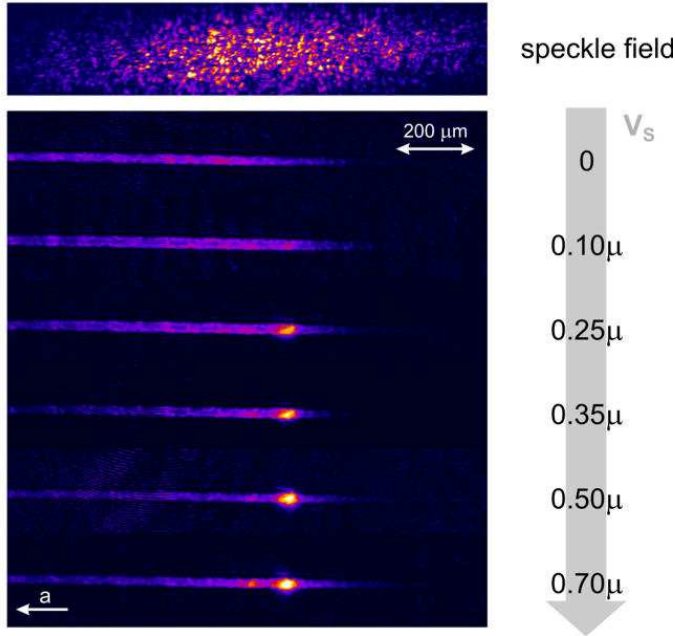}
\end{center}
\caption{Expansion of a BEC in a disordered optical guide. Top)
Intensity profile of the speckle field used in the experiment.
Bottom) Density profiles of the condensate after expansion in the
disordered optical guide for different speckle heights $V_S$, here
expressed in units of the BEC chemical potential $\mu=2.5$ kHz in
the initial trap. Adapted from \citet{fort05}.}
\label{fig:randomexp}
\end{figure}

\begin{figure}
\begin{center}
\includegraphics[width=0.8\columnwidth]{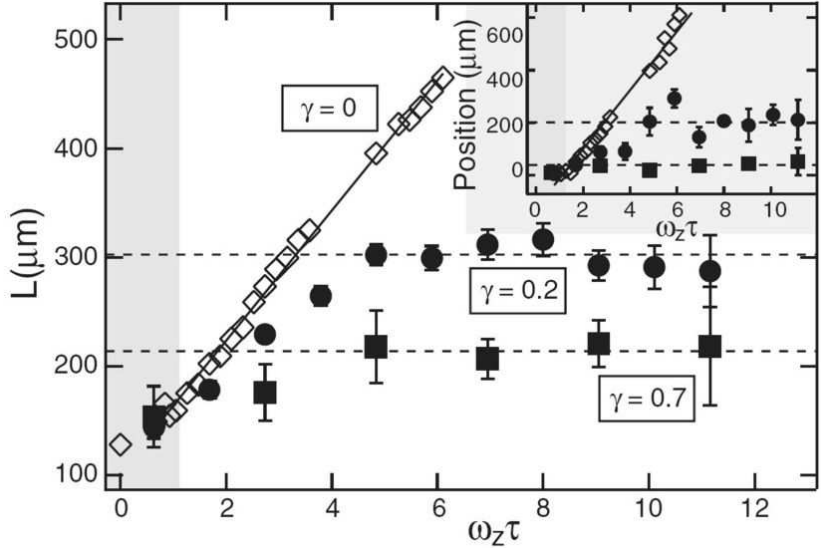}
\end{center}
\caption{Expansion of a BEC in a disordered magnetic waveguide. Time
evolution of the axial BEC rms size $L$ and center-of-mass position
(inset) for various amplitudes $\gamma$ of the random speckle
potential (in units of the BEC chemical potential). Taken from
\citet{clement05prl}.} \label{fig:clement}
\end{figure}

In Fig. \ref{fig:randomexp}, taken from \citet{fort05}, the density
distribution of the condensate, imaged \emph{in situ} after a fixed
expansion time in the optical waveguide, is shown for different
speckle potential heights (ranging from $V_S=0$ to $V_S=0.7 \mu$,
with $\mu$ the BEC chemical potential) together with the picture of
the actual speckle field used. Without speckles the condensate
freely expands, while in the presence of the speckles both the
expansion and the center-of-mass motion (induced by a small
acceleration along the waveguide) start to be suppressed for $V_S
\gtrsim 0.3 \mu$. A closer look shows that actually two different
components can be distinguished: while a low density cloud expands
without stopping, a few localized density peaks become observable
when increasing the speckle height. In Fig. \ref{fig:clement}, taken
from \citet{clement05prl}, the rms size of the BEC expanding in a
disordered magnetic waveguide is plotted as a function of time for
different heights of the disordered potential: one can clearly see
the transition from a diffusive regime in the absence of disorder to
a ``localization" regime when disorder is present.

Further investigations have demonstrated that this suppressed
expansion is not Anderson localization, but a classical localization
that can be explained with simple energetic arguments. The expanding
condensate is not a monochromatic flux of atoms all moving with the
same velocity: since the momentum distribution of the sample has a
finite width (mainly caused by the atom-atom repulsive interactions
which initially drive the expansion), a low velocity component of
the cloud is always present and get trapped in the speckles since it
has not sufficient energy to escape the deepest potential wells (in
the case of red-detuned speckles, as in \citet{fort05}) or to tunnel
through the highest potential barriers (in the case of blue-detuned
speckles, as in \citet{clement05prl}).

Several theoretical works have studied the expansion of an
interacting Bose-Einstein condensate in a speckle potential
\citep{clement05njp,modugno06,shapiro07,sanchezpalencia07,akkermans08,sanchezpalencia08}.
In particular, in \citet{clement05prl,clement05njp} and
\citet{modugno06} it has been shown that the expanded BEC density
profile is actually made up of two spatially separated parts. In the
center of the cloud interaction energy is dominating over kinetic
energy and the BEC density profile exactly follows the shape of the
potential, as expected from the Thomas-Fermi approximation for an
interacting Bose gas: no Anderson localization is expected to appear
in this region. In the wings of the cloud the density is much
smaller and kinetic energy is dominating over interaction energy:
here the BEC almost behaves as a noninteracting gas and the density
profile has deviations from the Thomas-Fermi approximation. However,
numerical studies based on the Gross-Pitaevskii equation evidenced
that no Anderson localization is present even in this region.

There are two possible physical reasons impeding the observation of
Anderson localization. The first is indeed the presence of
interactions: from an intuitive point of view, repulsive
interactions between the atoms force them to spread more in space,
contrasting localization. The second reason is the finite
correlation length of the disorder: even in the absence of disorder,
Anderson localization could not be observable because the disordered
potential is not ``good" enough to produce the scattering strength
which is necessary to have a localization length smaller than the
system size. It is well known that, in the pure random case, any
infinitesimal amount of disorder leads to localization in 1D
\citep{abrahams79}, but in finite-sized systems (like a trapped
Bose-Einstein condensate) the localization length plays an important
role.

The further experiments reported of \citet{fort05} have evidenced
that, apart from the problem of interactions, the typical speckle
potentials employed so far in the experiments were not
``fine-grained" enough to produce quantum reflection/transmission,
which is at the basis of 1D Anderson localization. This has been
observed by studying the collision of a BEC with a potential defect
created with a tightly focused laser beam, that mimics the effects
of one single speckle grain. The absence of quantum reflection from
the potential well created with this optical (red-detuned) defect
indicated that, in order to have quantum scattering, one should use
much steeper potentials, i.e. speckle potentials in which the
autocorrelation length $\sigma$ is smaller (the typical correlation
length of the speckles used in \citet{fort05} and
\citet{clement05prl} was 5 $\mu$m).

This problem has been theoretically addressed in \citet{modugno06},
where the coefficients of quantum reflection by a potential well and
of quantum transmission from a potential barrier have been
calculated as a function of the potential steepness and of the
velocity of the incident matter waves. Of course, quantum
reflection/transmission is fundamental in the case of 1D
localization, in which scattering just happens along a line: hence,
for multiple scattering to appear, only a fraction of the incident
wave has to be reflected/transmitted. In higher dimensions
interference due to multiple scattering could happen also for
classical reflection from potential hills, for which the
requirements are less stringent. However, in higher dimensions
localization itself is more difficult to achieve, owing to the
larger localization lengths (if localization is present).

Despite the obstacles discussed above, Anderson localization could
be eventually observed in 1D diffusion experiments similar to the
ones reported in \citet{clement05prl} and \citet{fort05}, provided
that the speckle autocorrelation length is made small enough. In
\citet{sanchezpalencia07} the density profile of the BEC expanding
in a weak speckle potential has been analytically worked out,
evidencing exponential localization in the dilute tails of the
wavefunction, where density is very low and interactions can be
neglected. In the case of interacting $^{87}$Rb BECs this could
happen for very small disorder correlation length, one order of
magnitude less than the ones achieved in \citet{clement05prl} and
\citet{fort05}. The crucial parameter, as we shall see in the
following section, is the ratio between the disorder correlation
length and the \emph{healing length}, which is the length scale
associated to the effect of interactions.
In \citet{sanchezpalencia07} a mobility edge was also found as a
maximum wavevector of the expanding BEC above which localization
cannot be observed, which is a peculiar characteristic of speckle
potentials with finite correlation length.

\subsubsection{Localization in a bichromatic lattice}
\label{sec:weaklybichromatic}

Quasiperiodic lattices, introduced in Sec. \ref{sec:bichromatic},
are a particular class of potentials which exhibit properties common
to both periodic and disordered systems \citep{diener01}. As in the
case of periodic lattices, their spectrum show reminiscence of
energy bands. On the other hand, owing to the lack of any
translational invariance, they support the existence of localized
states, which behave very similarly to the ones supported by truly
disordered systems \citep{grempel82}. Therefore they can be used as
a tool to study quantum localization, as a valid alternative to
speckle patterns, with the experimental advantage of an effortless
production of short length-scale potential fluctuations.

Localization in incommensurate bichromatic potentials is a well
known topic. This problem has been studied in detail in the
framework of the Harper model \citep{harper55} and of the 1D
tight-binding Aubry-Andr\'e model \citep{aubry80}, which is
described by the Hamiltonian
\begin{equation}
\hat{H} = - J \sum_n \left( |n\rangle \langle n+1| +
|n+1\rangle\langle n| \right)+ \Delta \sum_n \cos (2 \pi \beta n)
|n\rangle\langle n| \; , \label{eq:aubry}
\end{equation}
where $J$ is the tunneling rate between next-neighboring sites and
$\Delta$ is the amplitude of the quasiperiodic modulation of the
potential energy, being $\beta$ an irrational number. The
Aubry-Andr\'e model can be experimentally realized when the primary
lattice height $s_1$ is much larger than the secondary lattice
height $s_2$ (see Sec. \ref{sec:bichromatic}). The primary lattice
discretizes the system and produces a renormalization of the
effective mass $m^* = m E_{R1} / J \pi^2$, while the height of the
secondary lattice $s_2=\Delta / E_{R2}$ is the control parameter
which drives the localization transition. As a matter of fact,
differently from what happens with a pure disordered potential, in
the quasiperiodic case a localization transition exists even in 1D,
with a critical value $\Delta \approx 2 J$ of the quasidisorder
amplitude for producing a localized ground state.

\begin{figure}
\begin{center}
\includegraphics[width=0.9\columnwidth]{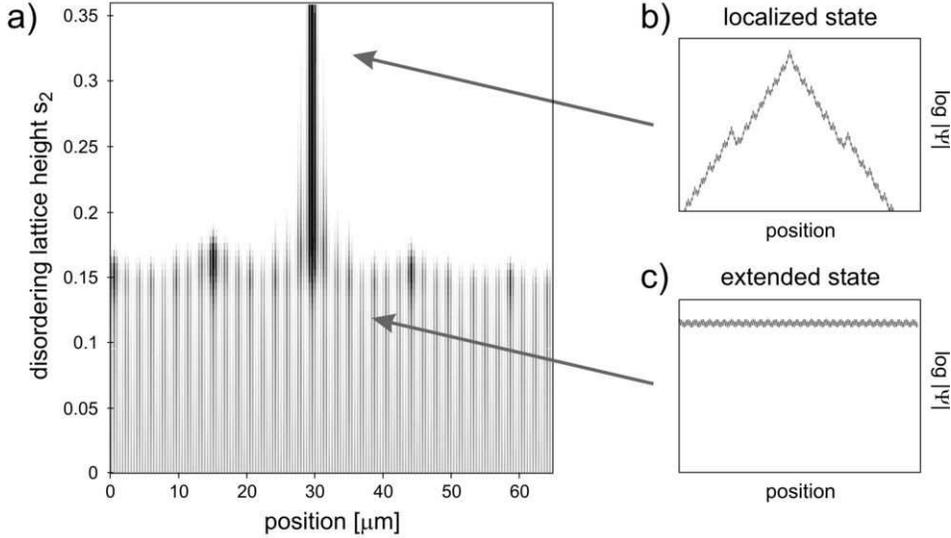}
\end{center}
\caption{Transition from extended to localized states in a
bichromatic incommensurate potential. a) The square modulus of the
ground state wavefunction is plotted in grayscale as a function of
position and disordering lattice strength. b,c) Logarithmic plots of
the ground state wavefunction below and above the localization
transition.} \label{fig:densityplotnew}
\end{figure}

This behavior is illustrated in Fig. \ref{fig:densityplotnew}a, in
which we plot the lowest-energy single-particle eigenstate in the
incommensurate lattice, as obtained by numerical integration of the
1D Schr\"{o}dinger equation (\ref{eq:se}), which holds in the
continuum. The figure shows, in grayscale, the squared modulus of
the ground state wavefunction as a function of position (horizontal
axis) and disordering lattice strength (vertical axis). One clearly
sees that for low values of disorder the ground state is an extended
state (see Fig. \ref{fig:densityplotnew}c), i.e. the wavefunction
extends across the entire lattice (in the limit $\Delta=0$ one
recovers the extended Bloch state describing the ground state in a
periodic lattice). Increasing disorder above a threshold value the
wavefunction suddenly localizes around few lattice sites, with
exponentially decreasing tails, in the same way as in the case of
Anderson localized states in $\delta$-correlated disordered
potentials (Fig. \ref{fig:densityplotnew}b).

\begin{figure}[t!]
\begin{center}
\includegraphics[width=0.7\columnwidth]{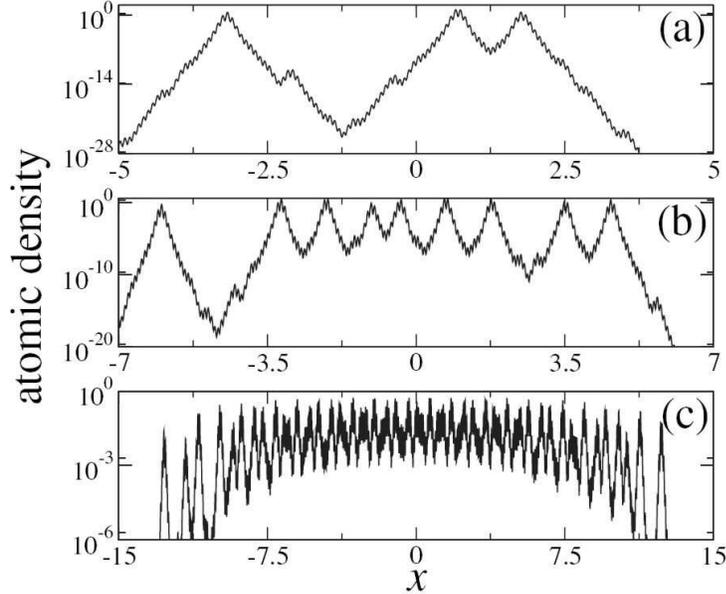}
\end{center}
\caption{Ground state of the Gross-Pitaevskii equation for a
Bose-Einstein condensate in a three-colour optical lattice for
different effective interaction strength $\tilde{g}=0.2$ (a),
$\tilde{g}=8$ (b), $\tilde{g}=256$ (c). The latter value corresponds
to the interaction strength for a Bose-condensed sample of $N=10^4$
$^{87}$Rb atoms in an elongated harmonic trap with frequencies
$\omega_\perp=2\pi \times 40$ Hz and $\omega_\parallel=2\pi \times
4$ Hz. Taken from \citet{schulte05}.} \label{fig:schulte}
\end{figure}

So far, we have shown that the incommensurate lattice supports the
existence of localized states for noninteracting particles. What
happens when one introduces interactions? This problem has been
discussed in \citet{schulte05,schulte06}, where the ground state of
the system has been calculated for a three-colour optical lattice.
Introducing repulsive interactions between the atoms, the numerical
integration of the 1D Gross-Pitaevskii equation shows that the
ground state wavefunction becomes a superposition of many
single-particle localized states, which add up to form an overall
extended state, as shown in Fig. \ref{fig:schulte} for different
interaction strengths. Similar results for a bichromatic lattice
have been presented in \citet{lye07}.

This behavior can be interpreted in terms of a \emph{screening
effect} induced by interactions \citep{sanchezpalencia06,schulte06}.
The nonlinear term in the Gross-Pitaevskii equation (\ref{eq:gpe})
can be treated as an effective potential cancelling the spectral
components of the original potential varying on length scales larger
than the \emph{healing length} $\xi=1/\sqrt{8\pi a n}$. This latter
quantity is the typical length scale that is associated to the
variation of the BEC wavefunction around sharp potential jumps
\citep{dalfovo99}. More generally, the healing length is the typical
length scale below which the condensate wavefunction is able to
behave quantum-mechanically. As a consequence, in order to observe
localization, the interference effects producing localization should
take place on a distance smaller than the healing length, otherwise
the BEC wavefunction would behave classically over longer
distancies. The healing length can be made larger by reducing the
amount of interactions in the system, that could be achieved either
by reducing the scattering length $a$ or by reducing the atom
density $n$. If the healing length is smaller than the disorder
localization length ($\xi < \zeta$, see Eq. \ref{eq:exploc}) no
localized states can be observed. If the healing length is larger
than the localization length, but smaller than the system size
($\zeta < \xi < L$), one could observe a superposition of many
localized states. Finally, if the healing length is the largest
length scale in the system ($\xi > L > \zeta$) the BEC wavefunction
collapses in a single localized state. This crossover is illustrated
in Fig. \ref{fig:schulte}, where the ground state in the three-color
lattice is plotted for the same lattice heights but different
interaction strength (hence same $\zeta$ but different $\xi$): while
in panel a) $\xi > \zeta$ and a few localized states can be easily
detected with clearly exponentially decreasing tails, in panel c)
$\xi < \zeta$ and an overall extended state forms.

\begin{figure}[t!]
\begin{center}
\includegraphics[width=1\columnwidth]{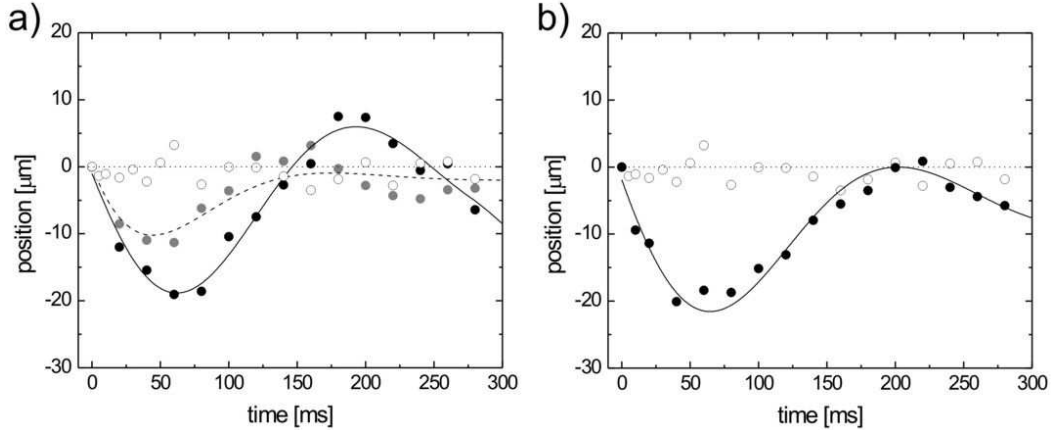}
\end{center}
\caption{Oscillations of a BEC in a parabolic + incommensurate
two-color lattice. a) Center-of-mass position for $N=1.5 \times
10^4$ atoms, $s_1=10$ and different disordering lattice heights
$s_2=0.1$ (filled black circles), $s_2=0.17$ (filled gray circles),
$s_2=0.25$ (empty circles). b) Center-of-mass position for $s_1=10$,
$s_2=0.25$ and different atom numbers $N=1.5 \times 10^4$ (empty
circles) and $N=2 \times 10^5$ (filled circles). Adapted from
\citet{lye07}.} \label{fig:lye}
\end{figure}

The existence of localized states can be probed with transport
experiments, similar to those presented in the previous section. In
\citet{lye07} the transport of an $^{87}$Rb BEC has been studied in
the presence of a bichromatic incommensurate potential. Localization
of the center-of-mass motion has been observed, the stronger the
smaller is the strength of interactions (tuned by changing the
atomic density), as shown in Fig. \ref{fig:lye}. This
density-dependent behavior, with interactions pushing to delocalize
the system, is reminiscent of Anderson-like localization. However,
in the regime of parameters studied in this work, no simple
Anderson-like localization has to be expected, being the eigenstates
of the system similar to the state dominated by interactions shown
in Fig. \ref{fig:schulte}c. The suppression of the center-of-mass
motion shown in Fig. \ref{fig:lye} was mainly caused by the strong
modulation of the BEC wavefunction on the length scale of the
beating between the two lattice periods, which resulted in very low
tunneling times across the lattice and, consequently, in an
extremely slow dynamics.

Effects of nonlinear dynamics have also been considered as possible
mechanisms to damp the motion. As a matter of fact, suppression of
transport is expected to appear for mechanisms alike the
interaction-induced dynamical instability observed in
\citet{cataliotti03}, \citet{fallani04} and \citet{cristiani04} for
monochromatic optical lattices. In the system studied in these works
the interplay between repulsive nonlinearities and band structure
resulted in fast-growing excitations dephasing the system and
halting the motion. The same effect can also be observed in
bichromatic optical lattices, in which a band structure can still be
identified \citep{diener01}, with a multitude of energy gaps opening
in the spectrum and getting denser and denser with increasing height
of the secondary incommensurate lattice.

From what discussed above we can draw a preliminary conclusion.
Bichromatic potentials do allow to solve the problem of the
correlation length of the speckles produced in
\citet{lye05,clement05prl,fort05,schulte05}, which causes the
localization length to be too large to be observable. In bichromatic
potentials the localization length can easily be smaller, however
the presence of too strong interactions still remains and makes it
impossible to observe a clear localization of the wavefunction in a
few localized states. In order to achieve clear signatures of
Anderson localization one needs to work with extremely weakly
interacting samples.

\subsection{Further directions}

\begin{figure}[t!]
\begin{center}
\includegraphics[width=0.6\columnwidth]{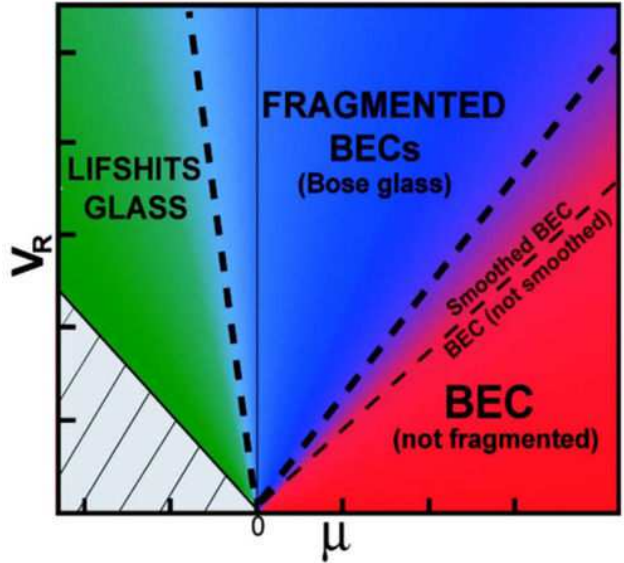}
\end{center}
\caption{Phase diagram of an interacting disordered BEC as a
function of the chemical potential $\mu$ and of the disordered
potential height $V_R$. Taken from \citet{lugan07}.}
\label{fig:lifshitz}
\end{figure}

From the theoretical point of view, as we have already pointed out,
the effect of disorder on the interacting Bose gas is an extremely
interesting topic of research. In \citet{lugan07} the ground state
of an interacting Bose gas in a disordered potential has been deeply
studied. In particular, a Lifshitz glass phase has been introduced
characterizing the ground state of the system for weak interactions.
Lifshitz states are a
particular class of localized states which exhibit ``weaker"
localization properties than Anderson-localized states, in the sense
that they show exponential decay only in the very far tails, while
close to the maximum their shape mostly depends on the \emph{local}
properties of the potential.
This means that they mostly resemble bound states of isolated
potential wells or trapped states between barriers, differently from
Anderson-localized states, whose shape is determined by
\emph{global} properties of the potential, i.e. by the combined
effect of many impurities / potential wells. In Fig.
\ref{fig:lifshitz} we show the phase diagram of the interacting
disordered BEC derived in \citet{lugan07} as a function of the BEC
chemical potential and of the speckle height. Starting from this
Lifshitz glass phase and increasing interactions, a phase of
fragmented interacting BECs has been proposed, which is a precursor
of the Bose glass phase (see Sec. \ref{sec:stronglyinteracting} for
further discussion).

The BEC fragmented state has been previously described in
\citet{wang04}. In this work the authors studied the ground state of
a BEC in the disordered potential produced by the random
imperfections of a magnetic microtrap. This paper was motivated by
several experimental observations
\citep{fortagh02,leanhardt03,esteve04,jones04}, in which
fragmentation of the BEC at very close distances from the
current-carrying wires of the microchip was observed. In the same
work \citep{wang04} the BEC dynamics in the disordered potential was
also investigated. In particular, the spectral analysis of the
sloshing motion after displacement of the confining potential
allowed to identify different dynamical regimes: superfluid
oscillations, self-trapping and an intermediate chaotic regime.

Recent theoretical works \citep{bilas06,lugan07bog} have also
studied the problem of Anderson localization of excitations in a
Bose-Einstein condensate. In the weakly interacting case, BEC
excitations are described by the Bogoliubov theory
\citep{dalfovo99}. By calculating the leading-order many-body
corrections to the classical BEC wavefunction, one finds that, in
the absence of external potentials, excitations are described by
quasiparticles with dispersion relation $\hbar \omega =\sqrt{ (\hbar
c k)^2 + (\hbar^2 k^2 / 2m)^2}$, with $k$ wavevector of the
excitation and $c$ sound velocity inside the BEC. This spectrum has
two distinct regions with different $k$-dependencies. For $k \ll
\xi^{-1}$ (with $\xi$ the healing length) the excitation spectrum is
phonon-like and excitations have energy $\omega \approx c k$. For $k
\gg \xi^{-1}$ the spectrum is particle-like and the energy of the
excitations is $\omega \approx \omega_0 + \hbar k^2/2m$, where
$\omega_0=mc^2/\hbar$ is an energy shift due to interactions.
\citet{bilas06}  have shown that, in the presence of a
white-noise random potential, excitations can undergo Anderson
localization, almost in the same way as the whole BEC wavefunction
can undergo. Further detailed studies have been carried out in
\citet{lugan07bog} in the more realistic case of correlated
disorder, showing that the localization length (and,
correspondingly, the possibility to observe localization in
finite-sized BECs) crucially depends on the correlation length of
the disorder.

The effect of disorder on the coherent BEC dynamics can be observed
also on the dephasing of Bloch oscillations which is expected to
appear when a disordered or quasi-disordered potential is
superimposed on a tilted optical lattice
\citep{sanchezpalencia05,schulte08}. Bloch oscillations are the
coherent oscillations of a wavepacket in a periodic potential when a
constant force is applied. This phenomenon has been observed for the
first time with ultracold atoms in optical lattices \citep{raizen97}
because of the much longer coherence times than the ones achievable
for electrons moving in real solid-state lattices, where defects and
impurities strongly dephase the system in a time much shorter than
the oscillation period. Also interactions lead to dephasing, as
evidenced in \citet{morsch01} when Bloch oscillations were observed
for the first time in a Bose-Einstein condensate. Later it was
demonstrated that this interaction-induced dephasing can be
controlled and eventually cancelled by tuning the interaction
strength with Feshbach resonances, as recently demonstrated in
\citet{gustavsson07} and \citet{fattori07}, or by using ultracold
fermionic samples \citep{roati04}, for which interactions are
forbidden by the Pauli principle. Noninteracting particles in
perfectly periodic optical lattices perform undamped Bloch
oscillations and, thanks to this possibility, they can be used as
microscopic probes for high-precision measurements of forces at
small distances \citep{carusotto05}. Starting from this ideal
situation and adding disorder on top of the periodic lattice, one
can quantitatively study the dephasing induced by disorder in a
controlled way, as first theoretically studied in \citet{schulte08}
and then recently investigated experimentally in
\citet{drenkelforth08}.

Generally speaking, disorder leads to the disruption of coherent
effects. Quite interestingly, however, under certain conditions
disorder can induce a spontaneous ordering of the system. This
effect, known as \emph{random-field induced order}, has been
originally studied in the context of classical spin models, which in
the presence of disorder may exhibit a magnetization higher than in
the ordered case \citep{wehr06,sende07}. Two-component Bose-Einstein
condensates in the presence of a random Raman coupling between the
two states can be used to study this class of effects, as recently
proposed in \citet{niederberger08}.

\subsection{Anderson localization: the state of the art}

Starting from the first experiments with speckle potentials
\citep{lye05,fort05,clement05prl,schulte05}, the quest for Anderson
localization in Bose-Einstein condensates has been a strongly active
direction of research. On one side experimental groups have focused
on producing disordered potentials on thinner length scales
\citep{clement05njp,demarco07}, in order to increase the amplitudes
of quantum scattering and decrease the attainable localization
lengths. On the other side, the challenge is to reduce atom-atom
interactions in order to make localization observable. Once this is
obtained, it will be even more interesting to study the effect of
adding a controlled amount of interactions. For this purpose it can
be strongly helpful to take advantage of Feshbach resonances to tune
the scattering length $a$, which is the key parameter defining the
strength of interactions $g=4\pi \hbar^2 a/m$ (see Sec.
\ref{sec:weaklyinteractintro}). In this perspective, the choice of
the element under investigation is crucial. The first experiments
performed with BECs in disordered potentials
\citep{lye05,fort05,clement05prl,schulte05} have focused on
$^{87}$Rb, which is a quite convenient element for the
implementation of cooling schemes, but has the disadvantage of
having a quite large scattering length $a\simeq 100 a_0$ (with $a_0$
the Bohr radius) and no favorable Feshbach resonances at convenient
magnetic fields \citep{marte02}. A much easier tuning of atom-atom
interactions could be provided by different elements, such as
$^{7}$Li (which has been already studied in combination with laser
speckles in \citet{chen07}) or $^{39}$K, studied by \citet{roati07}.

While this review was being completed (march 2008) two experiments
have succeeded in observing Anderson localization of coherent matter
waves, in the groups of A. Aspect in France and here at LENS. In the
French experiment \citep{billy08} a Bose-Einstein condensate is left
free to expand in a disordered waveguide produced by combining a
weakly focused laser beam with a 1D speckle potential. Differently
from the conceptually similar experiments \citet{clement05prl} and
\citet{fort05}, the analysis of the in-situ density profiles shows
clear indication of exponentially decreasing tails, which is a
signature of Anderson localization. This have been made possible by
a combination of several factors: the small atomic density in the
tails (necessary to reduce the counteracting effect of
interactions), the small speckle autocorrelation length (necessary
to have many quantum scattering events during the diffusion) and the
high detection sensitivity (allowing the observation of exponential
decay of the density). The observed localization is then
quantitatively compared with the theory developed in
\citet{sanchezpalencia07,sanchezpalencia08}.

In the experiment at LENS \citep{roati08}, Anderson localization has
been observed for a noninteracting $^{39}$K BEC in an incommensurate
bichromatic lattice, similar to that used in \citet{lye07}. Here the
strategy to exclude the effect of interactions is different: instead
of working with dilute samples, interactions are cancelled by tuning
a static magnetic field in proximity of a Feshbach resonance to set
the scattering length to zero. The noninteracting condensate in the
quasiperiodic potential thus realizes the noninteracting
tight-binding Aubry-Andr\'{e} model of Eq. (\ref{eq:aubry}), which
exhibits a transition from extended to localized states for
increasing disorder. The crossover between extended to localized
states is studied in detail by looking at the expansion of the BEC
and by studying spatial and momentum distribution of the states, all
of which result in agreement with the Aubry-Andr\'{e} predictions.


\section{Strongly interacting regime}
\label{sec:stronglyinteracting} \label{sec:stronglyinteractingintro}

In the previous section we have discussed the physics of disordered
weakly-interacting bosonic systems. The theoretical description of
this regime is provided by the semiclassical Gross-Pitaevskii Eq.
(\ref{eq:gpe}), which describes the propagation of nonlinear matter
waves. When interactions are strong, however, this mean-field
description is not capable to fully explain the behavior of the
system. A more appropriate description is provided by a full quantum
theory, taking into account quantum correlations between particles.
Also in this strongly interacting regime disorder may induce
localized quantum phases: these have a different nature from
Anderson localization since correlations between particles are
important, whereas Anderson localization is essentially a
single-particle effect.

Experimentally, a convenient way to enter the strongly interacting
regime is provided by the use of optical lattices (which we have
already introduced in Sec. \ref{sec:bichromatic}). In a deep optical
lattice the system becomes effectively stronger-interacting because
of the combined effect of the tighter squeezing of the atom
wavefunction in the potential wells (with a consequent increase of
the local density) and of the increase in the effective mass due to
the finite tunneling times across the potential barriers (which
makes the kinetic energy less important with respect to the
interaction energy).

For a system defined on a lattice, starting from the full many-body
Hamiltonian, one can derive a simplified zero-temperature model, in
the approximation that all the particles occupy the fundamental
vibrational state of the lattice sites. In this limit the quantum
state of an interacting gas of identical bosons in a lattice
potential is well described by the second quantization Bose-Hubbard
Hamiltonian \citep{fisher89,jaksch98}
\begin{equation}
\hat{H} = -J \sum_{\left<j,j'\right>} \hat{b}_j^\dagger \hat{b}_{j'}
+ \frac{U}{2} \sum_{j} \hat{n}_j \left( \hat{n}_j - 1 \right) +
\sum_{j} \epsilon_j \hat{n}_j \label{eq:bosehubbard}
\end{equation}
where $\hat{b}_j$ ($\hat{b}_j^\dagger$) is the annihilation
(creation) operator of one particle in the $j$-th site,
$\hat{n}_j=\hat{b}_j^\dagger \hat{b}_j$ is the number operator, and
$\left<j,j'\right>$ indicates the sum on nearest neighbors. Each of
the three terms on the right-hand-side of Eq. (\ref{eq:bosehubbard})
accounts for a different contribution to the total energy of the
system: $J$ is the \emph{hopping energy}, proportional to the
probability of quantum tunneling of a boson between neighboring
sites, $U$ is the on-site \emph{interaction energy}, arising from
atom-atom on-site short-range interactions (repulsive for $^{87}$Rb,
for which $U>0$) and giving a nonzero contribution only if more than
one particle occupies the same site, and $\epsilon_j \in
[-\Delta/2,\Delta/2]$ is a site-dependent energy accounting for
inhomogeneous external potentials superimposed on the lattice.

\begin{figure}
\begin{center}
\includegraphics[width=0.78\columnwidth]{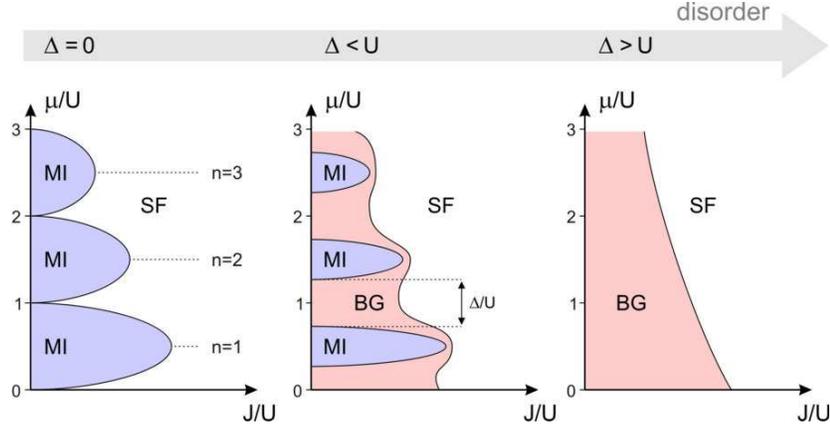}
\end{center}
\caption{Qualitative phase diagram for a disordered system of
lattice interacting bosons. Three phases can be identified: a
superfluid (SF), a Mott insulator (MI) and a Bose glass (BG).}
\label{fig:phasediagram}
\end{figure}

The quantum phase of the system depends on the interplay between
these three energy scales: hopping energy $J$, interaction energy
$U$ and disorder $\Delta$. We start considering the ideal case of a
translationally invariant system, in which $\Delta=0$. Assuming
integer filling of the sites, when $J>U$ the system is in a
superfluid (SF) state, in which the bosons are delocalized across
the lattice and the tunneling ensures off-diagonal long-range
coherence. Instead, when $U>J$, the system is in a localized Mott
insulator (MI) state, where long-range phase coherence is lost and
number Fock states are created at the lattice sites.  The actual
phase diagram of the system depends on the chemical potential
(related to the atomic density) and shows the existence of MI lobes
with integer number of atoms per site \citep{fisher89}. In the left
graph of Fig. \ref{fig:phasediagram} we show a qualitative sketch of
the phase diagram for a 3D system.

The transition from a SF to a MI for ultracold bosons in an optical
lattice has been proposed in \citet{jaksch98} and reported for the
first time in \citet{greiner02}, where the ratio $J/U$ was varied
across the transition point by controlling the height of the
lattice. The realization of a MI state does require a 3D optical
lattice, since, in order to enter the strongly interacting regime,
the atomic wavefunction should be squeezed in tightly confining
traps, with a site occupation on the order of unity. However, by
using deep optical lattices effectively slicing the atomic sample
into decoupled 2D or 1D systems, it is possible to study the SF-MI
transition in lower dimensionality, as made in \citet{stoferle04}
and \citet{spielman07}.

In the presence of a disordered external potential the additional
energy scale $\Delta$ enters the description of the system and is
responsible for the existence of a new quantum phase. In the
presence of weak disorder the MI lobes in the phase diagram should
progressively shrink and a new \emph{Bose glass} (BG) phase should
appear (central graph of Fig. \ref{fig:phasediagram}), eventually
washing away the MI region for $\Delta>U$ (right graph of Fig.
\ref{fig:phasediagram}) \citep{fisher89}. In a simplified view, a
Bose glass is half-way from a Mott insulator to a superfluid: it is
an insulating state, with no long-range phase coherence, as the Mott
insulator is; nevertheless, it is compressible and has no energy gap
in the excitation spectrum, as a superfluid has.

The Bose glass phase has been first identified in
\citet{giamarchi88}, where strongly interacting 1D bosonic systems
were studied. In the '90s it was widely studied in the context of
the superfluid-insulator transition observed in many
condensed-matter systems, such as $^{4}$He adsorbed on porous media
\citep{crowell95}, thin superconducting films \citep{goldman98},
arrays of Josephson junctions \citep{vanderzant92} and
high-temperature superconductors \citep{jiang94,budhani94}. The
possible realization of a Bose glass in a system of ultracold bosons
in a disordered lattice has been first proposed in
\citet{damski03,roth03}. More recently, the phase diagram of this
system has been derived in other theoretical papers, considering
also finite temperature effects \citep{krutitsky06,buonsante07pra},
detection schemes \citep{bargill06} and the possible realization of
a Bose glass with incommensurate bichromatic lattices
\citep{pugatch06,roscilde07}. Evidences for Bose glass phases have
been also theoretically obtained for different classes of
Bose-Hubbard models, where disorder is introduced either in the
hopping energy \citep{buonsante07} or in the on-site interaction
energy \citep{gimperlein05}.

The Bose glass is just the simplest disordered quantum phase that
can be realized in the strongly interacting regime. When atoms of
different species, or different internal (spin) states of the same
species, are considered, more complicated models can be
experimentally realized and new disordered quantum phases can
emerge. Atomic Bose/Fermi mixtures, in particular, represent a
versatile system in which many different disordered models can be
realized \citep{sanpera04,ahufinger05}. In the strong interacting
limit this system can be described in terms of composite fermionic
particles corresponding to one fermion + one bosonic particle/hole
in the same site. \citet{sanpera04} have shown that the interaction
between these composite fermions can be tuned by changing the
external potential: thus, a disordered potential can be used to
induce an effective random interaction between the particles. This
possibility allows the investigation of a variety of
disordered-related models, from fermionic Ising spin glasses to
models of quantum percolation \citep{ahufinger05}.

\subsection{The quest for Bose glass}

Experiments with disordered bosons in the strongly interacting
regime started at LENS in 2006. The system under investigation was a
collection of 1D atomic systems in a bichromatic optical lattice. A
main optical lattice was used to induce the transition from a weakly
interacting superfluid to a strongly correlated Mott insulator. A
secondary optical lattice was then used to add controlled
quasi-disorder to the perfect crystalline structure of the MI phase.
With reference to Eq.( \ref{eq:bosehubbard}), the non-commensurate
periodic potential superimposed on the main lattice introduces
inhomogeneities of the energy landscape $\epsilon_j \in
[-\Delta/2,\Delta/2]$ on the same length scale as the lattice
spacing.

\subsubsection{Excitation spectrum and coherence properties}

\begin{figure}
\begin{center}
\includegraphics[width=0.6\columnwidth]{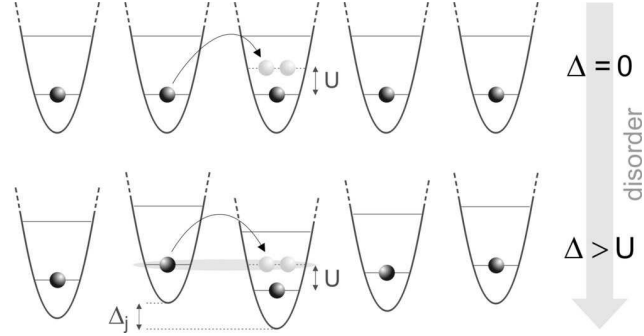}
\end{center}
\caption{Excitations in the deep insulating phases. a) In a Mott
insulator the tunneling of one boson from a site to a neighboring
one has an energy cost $\Delta E=U$. b) In the disordered case the
excitation energy is $\Delta E=U \pm \Delta_j$, that becomes a
function of the position. In the Bose glass state, in which
$|\Delta_j| > U$, an infinite system could be excited at arbitrarily
small energies and the energy gap would disappear.}
\label{fig:excitations}
\end{figure}

As introduced in the previous section, the excitation spectrum is an
important observable that can be measured in order to characterize
the quantum state of the system. By exploiting the possibility of
time-modulating the lattice potential, as first realized in
\citet{stoferle04}, it is possible to directly measure the
excitation spectrum and study how it is modified by the presence of
disorder. Naively speaking, in the MI phase one realizes a crystal
of atoms pinned at the lattice sites and sitting on the fundamental
vibrational level, as schematically shown in the top of Fig.
\ref{fig:excitations}. In a MI an energy gap in the excitation
spectrum exists, since the elementary excitation - the hopping of a
particle from a site to a neighboring one, or, in other words, the
creation of a particle-hole pair - has an energy cost $U$,
corresponding to the interaction energy of a pair of mutually
repelling atoms sitting on the same site (see Fig.
\ref{fig:excitations}).

\begin{figure}[t!]
\begin{center}
\includegraphics[width=\columnwidth]{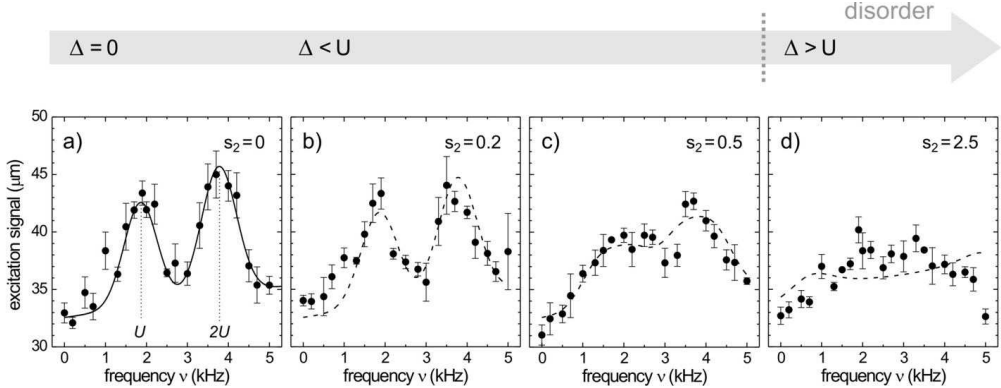}
\end{center}
\caption{Excitation spectra of the atomic system in a Mott insulator
state for increasing height of the disordering lattice. The
resonances are lost and the excitation spectrum becomes flat.
Adapted from \citet{fallani07}.} \label{fig:spectra}
\end{figure}

In Fig. \ref{fig:spectra}a we show the excitation spectrum of a Mott
insulator measured in the LENS experiments \citep{fallani07}. The
plot shows a well resolved resonance at energy $U$, which is
distinctive of the MI state, and a second resonance at energy $2U$.
While the physical origin of the excitation peak at $U$ is the
tunneling of particles between sites with the same occupancy, the
second peak at $2U$ can be ascribed to several processes: it can
arise from tunneling at the boundary between MI regions with
different site occupancy (that are present due to the inhomogeneity
of the confined sample), from higher-order processes and from
nonlinear effects due to the strong modulation. A theoretical
analysis of the response of the bosonic system to this lattice
modulation has been recently reported in \citet{kollath06} and
\citet{clark06}.

When increasing disorder the experiment showed a broadening of the
resonance peaks, which eventually become undistinguishable when
$\Delta \approx U$. As a matter of fact, the presence of disorder
introduces random energy differences $\Delta_j \in [-\Delta,\Delta]$
between neighboring sites (see bottom of Fig.
\ref{fig:excitations}). As a consequence, the tunneling of a boson
through a potential barrier costs $U \pm \Delta_j$, that becomes a
function of the position \citep{guarrera07}. The excitation energy
is not the same for all the bosons, differently from the pure MI
case, and the resonances become inhomogeneously broadened, as can be
observed in the experimental spectra at weak disorder ($\Delta<U$)
shown in Fig. \ref{fig:spectra}b,c \citep{fallani07}. This
broadening is in agreement with a semi-classical model
\citep{guarrera07} and has been recently predicted in theoretical
works \citep{hild06,zakrzewski08}, where the authors study the
dynamical response of a 1D bosonic gas in a superlattice potential
when a periodic amplitude modulation of the lattice is applied.

Eventually, when $\Delta \gtrsim U$, one expects that an infinite
system can be excited at arbitrarily small energies and that the
energy gap would shrink to zero. When this happens, nearby sites
become degenerate and regions of local superfluidity with
short-range coherence appear in the system. This novel many-body
state in which there is no gap but the system remains globally
insulating is a \emph{Bose glass}.

From the experimental point of view, additional information on the
nature of the many-body ground state can be acquired by analyzing
the density distribution of the atoms released from the lattice
after a time-of-flight. Long-range coherence in the sample results
in a density distribution with interference peaks at a distance
proportional to the lattice wavevector \citep{pedri01}. The
visibility of these peaks provides a measurement of phase coherence.
When increasing the height of the main lattice, a progressive loss
of long-range coherence has been reported in \citet{fallani07}
indicating the transition from a superfluid to an insulating state,
also in the presence of disorder. The combination of the excitation
spectra measurements and the time-of-flight images indicates that,
with increasing disorder, the system realized in \citet{fallani07}
goes from a MI to a state with vanishing long range coherence and a
flat density of excitations. The concurrence of these two properties
cannot be found in either a SF or an ordered MI, and is consistent
with the formation of a Bose glass, which is indeed expected to
appear for $\Delta \gtrsim U$.

Much work has still to be done for the exhaustive characterization
of such novel disordered state. New detection schemes should be
implemented, in order to have access to additional observables. This
necessity is not only restricted to the study of disordered systems,
being a more general issue shared by the experimental investigation
of different strongly interacting lattice systems, including e.g.
systems with magnetic ordering or mixtures of different species.
From the theoretical side, very recent works
\citep{roscilde07,roux08} have extensively studied the problem of 1D
interacting bosons in quasiperiodic lattices, working out the phase
diagrams (which include the presence of Bose glass and
incommensurate ``band insulating" / ``charge density wave" regions)
and studying how the different phases affect experimentally
detectable signals.

\subsubsection{Noise correlations}

\begin{figure}[t!]
\begin{center}
\includegraphics[width=0.8\columnwidth]{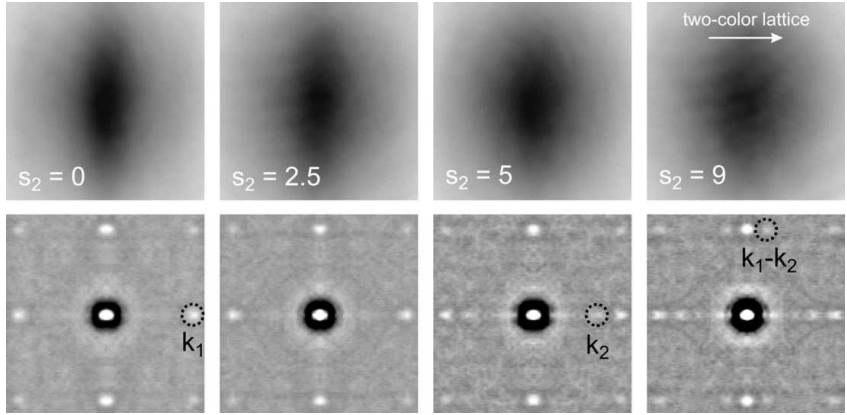}
\end{center}
\caption{Top) Time-of-flight absorption images of atoms in a Mott
insulator state for increasing height of the secondary lattice
$s_2$. Bottom) Density-density correlation functions corresponding
to the pictures above. The additional correlation peaks for large
$s_2$ arise from the destruction of the Mott domains and the
redistribution of the atoms in the lattice. Adapted from
\citet{guarrera08}.} \label{fig:noise}
\end{figure}

In the recent work \citet{guarrera08} noise interferometry has been
used to study interacting $^{87}$Rb bosons in the bichromatic
lattice. This detection technique, originally proposed in
\citet{altman04}, is based on the analysis of the spatial
density-density correlations of the atomic shot noise after
time-of-flight. These correlations are based on the Hanbury Brown \&
Twiss effect \citep{hanburybrown56}: if two identical particles are
released from two lattice sites, the joint probability of detecting
them in two separate positions (e.g. imaging them on two separate
pixels of a CCD camera) depends on the distance between the
detection points. These correlations, arising from quantum
interference between different detection paths, were first observed
for bosons in a Mott insulator state \citep{folling05} and then also
for band-insulating fermions \citep{rom06}. The sign of the
correlations depends on the quantum statistics: while bosons show
positive correlations (due to their tendency to bunch, i.e. to
arrive together at the detectors), fermions exhibit negative
correlations (due to the antibunching, consequence of the Pauli
exclusion principle). In the case of a bosonic Mott insulator, one
observes positive density-density correlation peaks at a distance
proportional to the lattice wavevector $k_1$ \citep{folling05}, as
shown in the first image of the bottom row of Fig. \ref{fig:noise}
for the recent experiment at LENS.

In \citet{guarrera08} noise correlations have been measured,
starting from a Mott insulator state, for increasing heights $s_2$
of the secondary lattice. The absorption images after time-of-flight
do not present significative differences, as shown in the top row of
Fig. \ref{fig:noise}, and demonstrate the absence of first order
(phase) coherence of the atomic system in the insulating state, even
in the presence of the secondary lattice. However, second order
(density) correlations turn out to be significantly different with
varying $s_2$, as illustrated in the noise correlation functions
plotted in the bottom row. More precisely, with increasing $s_2$,
one observes the appearance of additional correlation peaks at a
distance proportional to the wavevector $k_2$ of the secondary
lattice and to the beating between the two lattices $k_1-k_2$. These
peaks have to be associated with the redistribution of atoms in the
lattice sites as the disordering lattice is strengthened: the MI
regions characterized by uniform filling are destroyed and atoms
rearrange in the lattice giving rise to a state with non-uniform
site occupation, which follows the periodicity of the secondary
lattice. The redistribution of atoms is then quantitatively detected
by measuring the height of the additional correlation peaks.

Noise correlations thus prove to be a tool to extract important
information on the lattice site occupation, which is connected to
the second-order correlation function of the many-body state. The
appearance of similar correlation peaks was predicted in theoretical
works for hard-core bosons \citep{rey06} and soft-core bosons
\citep{roscilde07} in bichromatic lattices. Future works will study
the possibility to use noise interferometry to get additional
insight on the nature of the disordered insulating states produced
in the experiment, in particular in connection with the realization
of a Bose glass phase.

\subsection{Experiments with atomic mixtures}
\label{sec:atomicmixtures}

As we have discussed in Sec. \ref{sec:otherways}, disorder can be
produced by letting the atoms interact with randomly-distributed
scatterers of a different atomic species. The configuration proposed
in \citet{gavish05} has not yet been realized experimentally.
However, in 2006 the first experiments with binary mixtures in
optical lattices have been realized almost at the same time in two
different groups, in Zurich \citep{gunter06} and in Hamburg
\citep{ospelkaus06}. In these experiments $^{87}$Rb bosons and
$^{40}$K fermions were mixed together in a 3D optical lattice. Since
the two atoms have very similar resonance wavelengths, the depth of
the optical lattice is almost identical for the two species,
however, being potassium lighter than rubidium, its mobility is
favored. As a result, for a range of lattice heights, rubidium can
be localized in a Mott insulator state, while potassium atoms are
still able to move across the lattice.

In these experiments the superfluid to Mott insulator transition of
$^{87}$Rb was investigated as a function of the concentration of
$^{40}$K impurities, typically in the range 0 to 20\%. In
particular, the visibility of the interference pattern after
time-of-flight was investigated. The observation reported by the two
groups was a downshift of the lattice height value at which
coherence starts to be lost, when potassium atoms are introduced in
the system. Different interpretations for this effect have been
given, including finite-temperature effects, disorder-like induced
localization, or effects connected with the strong attractive
interaction between the two species.

Recently, a closely related system has been investigated at LENS
\citep{catani07} by using a binary bosonic mixture of $^{87}$Rb and
$^{41}$K in a 3D optical lattice. Similarly to the experiments
described above, loss of coherence in the rubidium sample induced by
the presence of potassium has been observed. However, this latter
experiment differs from the former ones in two points: the mixture
is bosonic/bosonic (instead of bosonic/fermionic) and the
interspecies interaction is repulsive (instead of attractive). The
observation of similar effects in systems with different quantum
statistics and different interaction sign rules out some of the
interpretation given so far, even if a clear explanation of the
observation has not yet been found. Future advances of these
experiments with mixtures in optical lattices will be pushed by the
use of Feshbach resonances for fine tuning of the interspecies
interaction.


\section{Conclusions}

The investigation of Bose-Einstein condensates in disordered
potentials is a fastly growing field of research. For decades
condensed-matter physicists have theoretically studied the interplay
between disorder and interactions in determining the transition from
metals to insulators. Now disordered systems can be realized in cold
atoms laboratories and, differently from traditional solid state
systems, they allow a fine tuning of both disorder and interactions,
as well as the advantage of new detection capabilities, thus
extending the range of the possible experimental investigations.

The first experiments were realized only a few years ago, with the
successful creation of disordered and quasi-disordered potentials
and the first studies of the behavior of ensembles of ultracold
bosons in different regimes of interactions. Although this field of
research is quite young, it already relies on an extensive
literature, mostly comprising theoretical works. The physics of
disordered atomic systems is indeed extremely rich, both in the
weakly interacting regime (where Anderson localization and its
disruption by interactions can be studied) and in the strongly
interacting regime (where particles are strongly correlated and new
quantum localized phases can emerge, as the Bose glass). The few
experiments performed until now have just opened a new direction,
showing the great potentialities of ultracold atoms for
investigating the physics of disorder, but still leaving open
questions concerning the observed localization effects. Regarding
this point, much more results are likely to come in the near future.

We have tried to give an overview of this newborn field of research,
discussing the topics of interest and the experimental efforts made
up to now. We would like to conclude by noting that any review is by
necessity incomplete and cannot be exhaustive of all the work made
in the field. For this reason we apologize with the authors of works
which we have unintentionally forgotten to mention.


\section{Acknowledgments}

This work has been supported by European projects SCALA and
EUROQUAM. We would like to thank D.~Cl\'ement and M.~Modugno for
careful reading of the manuscript and all the other members of the
Cold Quantum Gases group in Florence. We also acknowledge the
authors of the works described in this review who have kindly
granted the permission for using their figures.

\end{document}